\documentclass[reprint,NumberedRefs]{JASAnew}

\usepackage{natbib}
\usepackage{graphicx,color,rotating}
\usepackage[latin1]{inputenc}
\usepackage{textcomp}
\usepackage{dcolumn}
\usepackage{bm}     
\usepackage{upgreek}
\usepackage{subdepth}  			
\usepackage{epstopdf}   		
\usepackage[latin1]{inputenc}
\usepackage{amsmath}
\usepackage{amssymb}
\usepackage{amsfonts}
\usepackage{array}
\newcolumntype{L}[1]{>{\raggedright\let\newline\\\arraybackslash\hspace{0pt}}m{#1}}
\newcolumntype{C}[1]{>{\centering\let\newline\\\arraybackslash\hspace{0pt}}m{#1}}
\newcolumntype{R}[1]{>{\raggedleft\let\newline\\\arraybackslash\hspace{0pt}}m{#1}}

\usepackage{xcolor}						
\newcommand{\upspace}{\rule{0ex}{3.0ex}}

\newcommand{\mr}[1]{\ensuremath{\mathrm{#1}}}
\newcommand{\myvec}[1]{\bm{#1}}
\newcommand{\ee}{\mr{e}}
\newcommand{\ii}{\mr{i}}
\newcommand{\dm}{\mr{d}}

\newcommand{\avr}[1]{\big\langle{#1}\big\rangle}

\DeclareMathOperator{\re}{Re}

\newcommand{\iot}{{\ii\omega t}}

\newcommand{\ve}{\varepsilon}
\newcommand{\veO}{\ve_0}

\newcommand{\pp}{\partial}

\newcommand{\nablabf}{\boldsymbol{\nabla}}

\newcommand{\Lapl}{\nabla^2}

\newcommand{\pargrad}{\boldsymbol{\nabla}_\parallel}

\renewcommand{\etal}{\textit{et~al.}}

\newcommand{\scap}{\!\cdot\!}

\newcommand{\DDD}{\myvec{D}}

\newcommand{\een}{\myvec{e}}

\newcommand{\FFFrad}{\myvec{F}^\mr{rad}}

\newcommand{\Hsi}{{H_\mr{si}}}	
\newcommand{\Hpy}{{H_\mr{py}}}

\newcommand{\ks}{k_\mr{s}}

\newcommand{\nnn}{\myvec{n}}

\newcommand{\rrr}{\myvec{r}}

\newcommand{\uuu}{\myvec{u}}

\newcommand{\vvv}{\myvec{v}}

\newcommand{\vvvsl}{\vvv^{{}}_\mr{sl}}

\newcommand{\zerovec}{\boldsymbol{0}}

\newcommand{\calV}{\mathcal{V}}

\newcommand{\cfl}{c_\mr{fl}}
\newcommand{\cflsqr}{c^2_\mr{fl}}

\newcommand{\Eacfl}{E_\mr{ac}^\mr{fl}}

\newcommand{\kapPS}{\kappa_\mr{ps}}

\newcommand{\kapfl}{\kappa_\mr{fl}}

\newcommand{\Urad}{U^{\mr{rad}_{}}}

\newcommand{\etafl}{\eta_\mr{fl}}
\newcommand{\etaflb}{\eta^{{\mr{b}}}_\mr{fl}}

\newcommand{\Gamsl}{\Gamma_\mr{sl}}
\newcommand{\Gamfl}{\Gamma_\mr{fl}}

\newcommand{\rhofl}{\rho_\mr{fl}}
\newcommand{\rhosl}{\rho_\mr{sl}}

\newcommand{\fthf}{f_\mr{thf}}
\newcommand{\fpzt}{f_\mr{pzt}}

\newcommand{\pIconj}{p^{*_{}}_1}

\newcommand{\pII}{p_2}

\newcommand{\Vpp}{\text{V$_\mr{pp}$}}

\newcommand{\vvvI}{\vvv_1}

\newcommand{\vvvII}{\vvv_2}
\newcommand{\vvvIIbc}{\vvv^{{\mr{bc}}}_2}

\newcommand{\rhoPS}{\rho_\mr{ps}}

\newcommand{\SIC}{\textrm{C}}

\newcommand{\SICel}{^\circ\!\textrm{C}}

\newcommand{\SIum}{\upmu\textrm{m}}

\newcommand{\SIGHz}{\textrm{GHz}}
\newcommand{\SIMHz}{\textrm{MHz}}

\newcommand{\SIJpcm}{\textrm{J}\:\textrm{m$^{-3}$}}

\newcommand{\SIkg}{\textrm{kg}}
\newcommand{\SIkgm}{\textrm{kg}\:\textrm{m$^{-3}$}}

\newcommand{\SIm}{\textrm{m}}

\newcommand{\SImum}{\textrm{\textmu{}m}}

\newcommand{\SInm}{\textrm{nm}}

\newcommand{\SIPa}{\textrm{Pa}}
\newcommand{\SIkPa}{\textrm{kPa}}
\newcommand{\SIMPa}{\textrm{MPa}}

\newcommand{\SIpTPa}{\textrm{TPa}^{-1}}

\newcommand{\SImPas}{\textrm{mPa}\:\textrm{s}}

\newcommand{\SIs}{\textrm{s}}

\newcommand{\SImps}{\SIm\,\SIs^{-1}}

\newcommand{\SIV}{\textrm{V}}

\newcommand{\nn}{\nonumber}
\newcommand{\beq}[1]{\begin{equation} \eqlab{#1}}
\newcommand{\eeq}{\end{equation}}
\newcommand{\bsub}{\begin{subequations}}
\newcommand{\esub}{\end{subequations}}
\def\bal#1\eal{\begin{align}#1\end{align}}
\def\balat#1#2\ealat{\begin{alignat}{#1} #2 \end{alignat}}
\def\bsubal#1 #2\esubal{\bsuba{#1}\begin{align}#2\end{align} \esuba}     
\def\bsubalat#1#2#3\esubalat{\bsuba{#1} \begin{alignat}{#2} #3 \end{alignat} \esuba}
\newcommand{\bsuba}[1]{\bsub \eqlab{#1}}
\newcommand{\esuba}{\esub}

\newcommand{\eqlab}[1]{\label{eq:#1}}
\renewcommand{\eqref}[1]{Eq.~(\ref{eq:#1})}
\newcommand{\eqnoref}[1]{(\ref{eq:#1})}

\newcommand{\figref}[1]{Fig.~\ref{fig:#1}}
\newcommand{\fignoref}[1]{\ref{fig:#1}}

\newcommand{\figlab}[1]{\label{fig:#1}}

\newcommand{\secref}[1]{Section~\ref{sec:#1}}

\newcommand{\seclab}[1]{\label{sec:#1}}
\newcommand{\tabref}[1]{Table~\ref{tab:#1}}
\newcommand{\tabsref}[2]{Tables~\ref{tab:#1} and~\ref{tab:#2}}
\newcommand{\tablab}[1]{\label{tab:#1}}

\newcommand{\sigmabf}{\bm{\sigma}}

\newcommand{\sigmabfsl}{\bm{\sigma}^{{}}_\mr{sl}}

\newcommand{\cL}{c_\mr{lo}}
\newcommand{\cT}{c_\mr{tr}}

\newcommand{\Lwa}{L_\mr{wa}}
\newcommand{\Wwa}{W_\mr{wa}}
\newcommand{\Hwa}{H_\mr{wa}}
\newcommand{\Lgl}{L_\mr{gl}}
\newcommand{\Wgl}{W_\mr{gl}}
\newcommand{\Hgl}{H_\mr{gl}}
\newcommand{\Lsl}{L_\mr{sl}}
\newcommand{\Wsl}{W_\mr{sl}}
\newcommand{\Hsl}{H_\mr{sl}}
\newcommand{\Lthf}{L_\mr{thf}}
\newcommand{\Wthf}{W_\mr{thf}}
\newcommand{\Hthf}{H_\mr{thf}}
\newcommand{\Lpzt}{L_\mr{pzt}}
\newcommand{\Wpzt}{W_\mr{pzt}}
\newcommand{\Hpzt}{H_\mr{pzt}}

\definecolor{darkgreen}{rgb}{0.00, 0.50, 0.00}
\definecolor{DARKGREEN}{rgb}{0.00, 0.50, 0.00}
\definecolor{RED}{rgb}{1.00, 0.00, 0.00}
\definecolor{GREEN}{rgb}{0.00, 1.00, 0.00}
\definecolor{BLUE}{rgb}{0.00, 0.00, 1.00}
\definecolor{MAGENTA}{rgb}{1.00, 0.00, 1.00}

\newcommand{\AlScNx}{{Al$_{1\text{-}x}$Sc$_{x}$N}}
\newcommand{\AlScNiv}{{Al$_{0.6}$Sc$_{0.4}$N }}

\begin{document}

\title{Bulk acoustofluidic devices driven by thin-film transducers and whole-system resonance modes}

\author{Andr\'e G. Steckel}
\email{angust@fysik.dtu.dk}
\affiliation{Department of Physics, Technical University of Denmark,\\ DTU Physics Building 309, DK-2800 Kongens Lyngby, Denmark}

\author{Henrik Bruus}
\email{bruus@fysik.dtu.dk}
\affiliation{Department of Physics, Technical University of Denmark,\\
DTU Physics Building 309, DK-2800 Kongens Lyngby, Denmark}

\date{29 January 2021}

\begin{abstract}
In acoustofluidics, acoustic resonance modes for fluid and microparticle handling are traditionally excited by bulk piezoelectric transducers. In this work, we demonstrate by numerical simulations in three dimensions (3D) that integrated piezoelectric thin-film transducers constituting less than 0.1\% of the device work equally well. The simulations are done using a well-tested and experimentally validated numerical model. Our proof-of-concept example is a water-filled straight channel embedded in a mm-sized glass chip with a 1-$\SImum$ thick thin-film transducer made of \AlScNiv. We compute the acoustic energy, streaming, and radiation force, and show that it is comparable to that of a conventional silicon-glass device actuated by a bulk PZT transducer. The ability of the thin-film transducer to create the desired acoustofluidic effects in bulk acoustofluidic devices rely on three physical aspects: The in-plane-expansion of the thin-film transducer under the orthogonal applied electric field, the acoustic whole-system resonance of the device, and the high Q-factor of the elastic solid constituting the bulk part of the device. Consequently, the thin-film device is surprisingly insensitive to the Q-factor and resonance properties of the thin-film transducer.
\end{abstract}
\maketitle

\section{Introduction}
\seclab{intro}

An increasing number of microscale ultrasound acoustofluidic devices are used for applications within clinical diagnostics, biology, and forensic sciences.\cite{Lenshof2012a, Gedge2012, Sackmann2014, Laurell2014, Antfolk2017} Examples include but are not limited to rapid sepsis diagnostics by detection of bacteria in blood,\cite{Ohlsson2016} enrichment of prostate cancer cells in blood,\cite{Augustsson2012} high-throughput cytometry and multiple-cell handling,\cite{Zmijan2015, Ohlin2015} cell synchronization,\cite{Thevoz2010} single-cell patterning and manipulation,\cite{Collins2015, Guo2016} and size-independent sorting of cells.\cite{Augustsson2016} Furthermore, acoustofluidics has been used for massively parallel force microscopy on biomolecules,\cite{Sitters2015} acoustic tweezing,\cite{Drinkwater2016, Collins2016, Lim2016, Baresch2016} and non-contact microfluidic trapping and particle enrichment\cite{Hammarstrom2014a}

Most applications rely on one of two basic methods for exciting the ultrasound field. One method is based on surface acoustic waves, excited by interdigitated metallic  electrodes positioned on the surface of a piezoelectric (PZE) substrate. The other method relies on bulk acoustic waves excited locally in liquid-filled acoustic microchannels or microcavities defined in acoustically hard materials by an attached bulk transducer,\citep{Lenshof2012} or in cavities with a thin silicon-membrane lid driven by a
lead-zirconate-titanate (PZT) thin film.\cite{Reichert2018}

Recently, the bulk-acoustic-wave method have been extended the concept of whole-system ultrasound resonances (WSUR), where the resonant acoustic waves are defined by the whole system, and not just the microcavities.\citep{Moiseyenko2019, Bode2020} In this paper we extend the WSUR method by substituting the large bulk PZE transducer by a tiny PZE thin-film transducer integrated on the surface of the device and constituting less than 0.1\% v/v of the resulting device.

Integrated thin-film PZE transducers have been used extensively for actuating electromechanical systems, often made of aluminum nitride (AlN). Thin-film transducers made of AlN are structurally and chemically stable, they have a low dielectric and mechanical loss, they are compatible with standard silicon-based CMOS microfabrication techniques. Academic applications of AlN thin-film transducers include RF filters,\cite{Dubois1999} suspended microchannel resonators,\cite{DePastina2018} contour mode resonators,\cite{Piazza2006} switches,\cite{Zaghloul2014, Sinha2009} and accelerometers.\cite{Olsson2009} AlN thin-film have been deposited on substrates of sapphire, crystal quartz, fused silica, and silicon,\cite{Zhang2006} and on 30-$\SImum$-thick Si membranes.\cite{Masson2007, Fujikura2008} Commercially, AlN-sputtered thin films are used in thin-film bulk-wave acoustic resonator filters.\cite{Ruby2017} However, hitherto thin-film transducers have not yet been applied in MHz bulk acoustofluidic devices.

In this paper, based on a  well-tested and experimentally validated numerical model,\citep{Skov2019, Skov2019b, Bode2020} we demonstrate by three dimensional (3D) numerical simulations that glass chips with integrated PZE thin-film transducers constituting less than 0.1\% v/v of the system, form devices with an acoustofluidic response fully on par with that obtained in a conventional silicon-glass device actuated by a bulk PZT transducer. This, perhaps surprising, result offers several advantages for the practical application of thin-film transducers within acoustofluidics: Thin-film devices do not depend on resonance properties of the thin-film transducer itself, and thin-film devices can be fabricated by well-controlled reproducible microfabrication techniques subject to parallel mass-fabrication processes.

The paper is organized as follows: In \secref{Theory} we summarize the theory and numerical model used throughout the paper. In \secref{ThinfilmVSbulk} we present a proof-of-concept example showing that a thin-film acoustofluidic device perform on par with a conventional bulk-transducer device. In \secref{PhysicalAspects}
we discuss the physical principle of the thin-film transduction process, the robustness of the device to material, thickness, and quality factor of the thin-film transducer, the role of the shape of the thin-film-transducer electrodes, and the sensitivity of the device to shifts in the channel position away from exact centering in the glass chip. Finally, in \secref{Conclusion} we present our conclusions.

\section{Theory}
\seclab{Theory}

We use, and in this section briefly summarize, the theory and numerical model developed by Skov \etal\citep{Skov2019} including the effective boundary layer theory by Bach and Bruus.\citep{Bach2018} This model is well-tested and experimentally validated,\citep{Skov2019, Skov2019b, Bode2020} and although originally stated for acoustofluidic devices with bulk PZT transducers, it is trivially extended to describe other types of PZE transducers, including thin-film transducers.

\subsection{Governing equations for the time-harmonic fields}
\seclab{GovEquTimeharm}

We consider a time-harmonic electric potential $\tilde{\varphi}(\rrr,t)$, which excites the PZE transducer and induces a displacement field $\tilde{\uuu}_1(\rrr,t)$ in the solids as well as an acoustic pressure $\tilde{p}^{{}}_1(\rrr,t)$ in the fluid channel and in the coupling layer. All of these fields $\tilde{F}(\rrr,t)$ separates into a complex-valued amplitude $F(\rrr)$ and a complex time-harmonic phase factor with frequency $f$,
 \beq{harmonic_fields}
 \tilde{F}(\rrr,t) = F(\rrr)\:\ee^{-\iot},\; \text{ with }\; \omega = 2\pi f.
 \eeq
The phase factor $\ee^{-\iot}$ cancels out in the following linear governing equations, leaving just the amplitude fields.

From first-order perturbation theory follows that the acoustic pressure $p_1$ in the fluid is governed by the Helmholtz equation with damping coefficient $\Gamfl$,
 \beq{EquMotionFluid}
 \Lapl\ p_1 = -\frac{\omega^2}{\cflsqr} \big(1+\ii\Gamfl\big)\, p_1,
 \text{ with }
 \Gamfl=\Big(\frac{4}{3}\etafl+\etaflb\Big)\,\omega\kapfl,
 \eeq
where $\cfl$ is the speed of sound, $\rhofl$ is the density, $\kapfl = (\rhofl\cflsqr)^{-1}$ is the isentropic compressibility, and $\etafl$ and $\etaflb$ are the dynamic and bulk viscosity of the fluid, respectively, see \tabref{param_water}. The acoustic velocity $\vvv_\textrm{1,fl}$ of the fluid is proportional to the  gradient of the pressure $p_1$,
 \beq{VelocityPotential}
 \vvv_1 = -\ii\:\frac{1-\ii\Gamfl}{\omega\rhofl}\:\nablabf p_1
 \eeq

\begin{table}[t]
\centering
\caption{\tablab{param_water} Parameter values of water at $25~\SICel$ used in the numerical simulations.\citep{Muller2014}}
\begin{ruledtabular}
\begin{tabular}{ccccc}
 Parameter&  Value &$\qquad$ & Parameter & Value \\ \hline
 $\rhofl$ & $ 997~\SIkgm$ &  & $\etaflb$ & $2.485~\SImPas$ \\
 $\cfl$   & $1497~\SImps$ &  & $\Gamfl$  & $10.3~\mr{THz}^{-1} f$ \\
 $\kapfl$ & $448~\SIpTPa$ &  & $\etafl$	 & $0.890~\SImPas$ \\
\end{tabular}
\end{ruledtabular}
\end{table}

For the linear PZE transducer, whether made of AlN, \AlScNx, or PZT, the electrical potential $\varphi$ inside the PZT, is governed by Gauss's law for a linear, homogeneous dielectric with a zero density of free charges,
 \beq{GaussLaw}
 \nablabf \cdot \DDD = \nablabf \cdot \big[-(1+\ii\Gamma_\ve)\myvec{\varepsilon}\cdot \nablabf \varphi\big] = 0,
 \eeq
where $\DDD$ is the electric displacement field and $\myvec{\varepsilon}$ the dielectric tensor. The governing equation for the mechanical displacement field $\uuu_1$ in a linear elastic solid (including the PZE) with density $\rhosl$, is Cauchy's equation
 \beq{CauchyEq}
 -(1+\ii\Gamsl)\rhosl\omega^2\:\uuu_1 = \nablabf \cdot \sigmabfsl,
 \eeq
In the PZE, the complete linear electromechanical coupling relating the stress and the electric displacement to the strain and the electric field is given in Voigt notation as,
 \beq{StressStrainPiezo}
 \resizebox{\columnwidth}{!}{$
 \left( \begin{array}{c}
 \sigma^{{}}_{xx} \\  \sigma^{{}}_{yy} \\  \sigma^{{}}_{zz} \\ \hline
 \sigma^{{}}_{yz} \\  \sigma^{{}}_{xz} \\  \sigma^{{}}_{xy} \\ \hline
 D^{{}}_x \\ D^{{}}_y \\ D^{{}}_z \\
 \end{array}  \right)
 =
 \left( \begin{array}{c@{\:}c@{\:}c@{\:}|c@{\:}c@{\:}c@{\:}|c@{\:}c@{\:}c}
 C^{{}}_{11} & C^{{}}_{12} & C^{{}}_{13} & 0 & 0 & 0 & 0 & 0 & -e^{{}}_{31} \\
 C^{{}}_{12} & C^{{}}_{11} & C^{{}}_{13} & 0 & 0 & 0 & 0 & 0 & -e^{{}}_{31} \\
 C^{{}}_{13} & C^{{}}_{13} & C^{{}}_{33} & 0 & 0 & 0 & 0 & 0 & -e^{{}}_{33} \\ \hline
 0 & 0 & 0 & \!c^{{}}_{44} & 0 & 0 & 0 &  -e^{{}}_{15} & 0 \\
 0 & 0 & 0 & 0 & c^{{}}_{44} & 0 & -e^{{}}_{15} & 0 & 0 \\
 0 & 0 & 0 & 0 & 0 & c^{{}}_{66}  & 0 & 0 & 0 \\ \hline
 0 & 0 & 0 & 0 &  e^{{}}_{15} & 0 & \ve^{{}}_{11} &  0 & 0 \\
 0 & 0 & 0 &  e^{{}}_{15} & 0 & 0 & 0 & \ve^{{}}_{11} & 0 \\
 e^{{}}_{31} & e^{{}}_{31} & e^{{}}_{33} & 0 & 0 & 0  & 0 & 0 & \ve^{{}}_{33}\\
 \end{array}  \right) \;
 \left(   \begin{array}{c}
 \pp^{{}}_x u^{{}}_x \\ \pp^{{}}_y u^{{}}_y \\ \pp^{{}}_z u^{{}}_z \\ \hline
 \pp^{{}}_y u^{{}}_z +\!  \pp^{{}}_z u^{{}}_y \\ \pp^{{}}_x u^{{}}_z +\! \pp^{{}}_z u^{{}}_x  \\ \pp^{{}}_x u^{{}}_y +\!  \pp^{{}}_y u^{{}}_x \\ \hline
 -\pp^{{}}_x \varphi \\ -\pp^{{}}_y \varphi \\ -\pp^{{}}_z \varphi \\
 \end{array}   \right).$}
 \eeq
The remaining three components of the stress tensor are given by the symmetry relation $\sigma_{ik} = \sigma_{ki}$. Similarly, the Cauchy equation~\eqnoref{CauchyEq} governs $\uuu_1$ in a purely elastic solid, but now the stress-strain relation~\eqnoref{StressStrainPiezo} includes only the first six rows and first six columns, as $\DDD$ and $\varphi$ do not couple to $\sigmabfsl$ and $\uuu_1$. The parameter values are listed in \tabref{param_solids}.

\begin{table}[t]
\centering
\caption{\tablab{param_solids} Parameters of the solids at $25~\SICel$ used in the numerical simulation. For glass $C_{12} = C_{11} - 2C_{44}$. For all PZE used in this work $C_{12} = C_{11} - 2C_{66}$.}
\begin{ruledtabular}
{\footnotesize
\begin{tabular}{ccccc}
 Parameter &  Value & & Parameter  & Value \\ \hline
 \multicolumn{5}{l}{\emph{Thin-film aluminum nitride}, AlN \cite{Caro2015, Iqbal2018, Olsson2020}} \upspace \\
 $\rhosl $    & 3300 $\SIkg\:\SIm^{-3}$ &   & $\Gamsl$ & 0.0005 \\
 $C_{11}$   & 410.2 GPa  &  & $C_{33}$	& 385.0 GPa\\
 $C_{12}$   & 142.4 GPa  &  & $C_{44}$	& 122.9 GPa\\
 $C_{13}$   & 110.1 GPa  &  & $C_{66}$	& 133.9 GPa\\
 $e_{31}$   &$-1.0475~\SIC\:\SIm^{-2}$    &
 & $e_{15}$	& $-0.39~\SIC\:\SIm^{-2}$ \\
 $e_{33}$   & 1.46 $\SIC\:\SIm^{-2}$      &
 &  $\Gamma_\ve$ & $0.0005$ \\
 $\epsilon_{11}$   &  9 $\epsilon_{0}$ &
 & $\epsilon_{33}$ & 11 $\epsilon_{0}$
 \\[1mm]
 \multicolumn{5}{l}{\emph{Thin-film aluminum scandium nitride}, \AlScNiv \cite{Caro2015, Olsson2020}}   \\
 $\rhosl $    & 3300 $\SIkg\:\SIm^{-3}$ &   & $\Gamsl$ & 0.0005 \\
 $C_{11}$   & 313.8 GPa  &  & $C_{33}$	& 197.1 GPa\\
 $C_{12}$   & 150.0 GPa  &  & $C_{44}$	& 108.6 GPa\\
 $C_{13}$   & 139.2 GPa  &  & $C_{66}$	& 81.9 GPa\\
 $e_{31}$   &$-2.65~\SIC\:\SIm^{-2}$    &
 & $e_{15}$	& $-0.32~\SIC\:\SIm^{-2}$ \\
 $e_{33}$   & 2.73 $\SIC\:\SIm^{-2}$      &
 &  $\Gamma_\ve$ & $0.0005$ \\
 $\epsilon_{11}$   &  22 $\epsilon_{0}$ &
 & $\epsilon_{33}$ & 22 $\epsilon_{0}$
 \\[1mm]
 \multicolumn{5}{l}{\emph{Bulk and thin-film lead zirconium titanate}, PZT~\cite{Ferroperm2017}}  \\
 $\rhosl $    & 7700 $\SIkg\:\SIm^{-3}$ &   & $\Gamsl$ & 0.005 \\
 $C_{11}$   & 168 GPa   &  & $C_{33}$	&  123 GPa\\
 $C_{12}$   & 110 GPa   &  & $C_{44}$	& 30.1 GPa\\
 $C_{13}$   & 99.9 GPa  &  & $C_{66}$	& 29.0 GPa\\
 $e_{31}$   &$-2.8~\SIC\:\SIm^{-2}$    &
 & $e_{15}$	& $9.86~\SIC\:\SIm^{-2}$ \\
 $e_{33}$   & 14.7 $\SIC\:\SIm^{-2}$      &
 &  $\Gamma_\ve$ & $0.005$ \\
 $\ve_{11}$ & 828 $\veO$ &  & $\ve_{33}$ & 700 $\veO$
 \\[1mm]
 \multicolumn{5}{l}{\emph{Glass, Schott D263}  \cite{SchottD263}} \\
 $\rhosl$ & 2510 $\SIkgm$ &  && \\
 $E$      & 72.9 GPa      &  & $s$  & 0.208           \\
 $C_{11}$ & 81.8 GPa      &  & $C_{44}$    & 30.2 GPa        \\
 $C_{12}$ & 21.5 GPa      &  & $\Gamsl$  &  0.0004  \\
 $\cL$    & 5710 $\SImps$ &  & $\cT$    & 3467 $\SImps$
 \\[1mm]
 \multicolumn{5}{l}{\emph{Glass, Pyrex}  \cite{CorningPyrex, Hahn2015}} \\
 $\rhosl$ & 2230 $\SIkgm$ &  && \\
 $E$      & 62.8 GPa      &  & $s$  & 0.20           \\
 $C_{11}$ & 69.7 GPa      &  & $C_{44}$    & 26.1 GPa        \\
 $C_{12}$ & 17.4 GPa      &  & $\Gamsl$  & 0.0004  \\
 $\cL$    & 5594 $\SImps$ &  & $\cT$    & 3425 $\SImps$
 \\[1mm]
 \multicolumn{5}{l}{\emph{Silicon substrate} \cite{Hopcroft2010, Hahn2015}} \\
 $\rhosl $  & 2329 $\SIkgm$ &   & $\Gamsl$ & 0.0001 \\
 $C_{11}$   & 165.7 GPa  &  & $C_{44}$	&  79.6 GPa\\
 $C_{12}$   & 63.9 GPa  &  &  	& \\
\end{tabular}
}
\end{ruledtabular}
\end{table}

\subsection{Governing equations for the steady time-averaged fields}
\seclab{GovEquTimeavr}

The non-linearity of the governing equation results in higher order responses to the time-harmonic actuation. Here, we are only interested in the steady time-averaged second-order response and define $F_2(\rrr) =\avr{F_2(\rrr,t)} = \frac{\omega}{2\pi} \int_0^{\frac{2\pi}{\omega}} F_2(\rrr,t)\, \dm t$. A time-average of a product of two first-order fields is also a second-order term, written as $\avr{A_1 B_1}=\frac{1}{2} \re \big[ A_1 B_1^*\big]$, where the asterisk denote complex conjugation. The acoustic streaming $\vvv_2$ is such a time-averaged field. It is a steady-state, incompressible Stokes flow driven by the slip velocity stated in \secref{BC} and the time-averaged acoustic dissipation body force proportional to $\Gamfl$,\citep{Bach2018}
 \beq{v2Stokes}
 \etafl\Lapl\vvvII = \nablabf \pII - \frac{\Gamfl\omega}{2\cflsqr}\:\re\Big[\pIconj\vvvI\Big],
 \qquad \nablabf\cdot\vvvII = 0.
 \eeq

The time-averaged acoustic energy density in the fluid is given as the sum of the kinetic and compressional energy densities,
 \beq{EacDef}
 \Eacfl = \frac14 \rhofl \big| \vvv_\mr{1,fl}\big|^2 + \frac14 \kapfl \big|p_\mr{1,fl}\big|^2.
 \eeq

The acoustic radiation force $\FFFrad$ acting on particles in the fluid is the gradient of the potential $\Urad$, specified for particles with radius $a$, density $\rhoPS$, and compressibility $\kapPS$, suspended in a fluid with density $\rhofl$ and compressiblity $\kapfl$,\cite{Settnes2012}
 \bsubal{FradEq}
 \FFFrad &= -\nablabf \Urad,
 \\
 \Urad &=
 \pi a^3\Big(\frac13 f_0\: \kapfl  |p_\mr{1,fl}|^2
 - \frac12 f_1\:\rhofl |\vvv_\mr{1,fl}|^2 \Big), \\
 f_0 &= 1 - \frac{\kapPS}{\kapfl}, \qquad
 f_1 = \frac{2(\rhoPS-\rhofl)}{2\rhoPS + \rhofl},
 \esubal
where, $f_0$ and $f_1$ is the so-called acoustic monopole and dipole scattering coefficient, respectively.

\subsection{Boundary conditions fluids, solids, and PZE}
\seclab{BC}

The boundary conditions of the fields on all boundaries and interfaces of the model are specified as follows. On the surfaces facing the surrounding air, we assume zero stress on the solid and the PZE as well as zero free surface charge density on the PZE. On the surfaces with electrodes, the PZE has a specified ac-voltage amplitude. On the internal surfaces between solid and PZE, the stress and displacement are continuous, and likewise on the fluid-solid interface, but here for the latter in the form of the effective  boundary conditions derived by Bach and Bruus.\citep{Bach2018}  These effective boundary conditions include the viscous boundary layer analytically, and thus we avoid resolving these very shallow boundary layers numerically. The effective boundary conditions include the velocity $\vvvsl = -\ii \omega \uuu_1$ of the solid (sl) and the complex-valued shear-wave number $\ks = (1+\ii)\:\delta_\textrm{fl}^{-1}$ of the fluid (fl), where $\delta_\textrm{fl}=\sqrt{2\etafl/(\rhofl\omega)} \approx 0.5~\SImum$ is the thickness of the boundary layer.

 \bsubalat{bcAll}{2}
 \eqlab{bcPZTphi}
 & \text{sl-fl:}
 & \sigmabf_\textrm{sl} \cdot \nnn & = - p_\textrm{1,fl}\: \nnn + \ii\ks\etafl(\vvvsl - \vvv_\textrm{1,fl}\big),
 \\
 & \text{sl-air:}
 & \sigmabf_\textrm{sl} \cdot \nnn &= \zerovec,
 \\
 & \text{PZE-air:}
 &\DDD\cdot\nnn &= 0,
 \\
 & \text{top elec.:}
 & \varphi &= 0,
 \\
 & \text{bot elec.:}
 & \varphi &= \pm \tfrac12 \varphi_0.
 \esubalat
Further, at fluid-solid interfaces, the slip velocity $\vvvIIbc$ driving the streaming takes into account both the motion of the surrounding elastic solid and the Reynolds stress induced in viscous boundary layer in the fluid,\citep{Bach2018}
 \bsubal{streaming}
 \eqlab{v2bc}
 \vvvII &= \vvvIIbc
 \qquad
 \nnn\cdot\vvvIIbc  = 0,
 \\ \nn
 (\bm{1}-\nnn\nnn)\cdot\vvvIIbc & =
 -\re\!\bigg[\!\bigg(\!\frac{2-\ii}{4\omega}\pargrad\scap\vvv_{1\parallel}^{*}
 \!+\! \frac{\ii}{2\omega}\pp^{{}}_\perp v^{*}_{1\perp}\bigg)\vvv_{1\parallel}\!\bigg]
 \\
 & \quad -\frac{1}{8\omega}\pargrad\big|\vvv_{1\parallel} \big |^2 .
 \esubal
Here, we have stated the special case of the slip velocity $\vvv^\mr{bc}_{2}$, which is only valid near acoustic resonance, where the magnitude $|\vvvI|$ of the acoustic velocity in the bulk is much larger than $\omega\:|\uuu^\mr{bc}_\mr{1,sl}|$ of the walls.

Finally, we use the symmetry at the $yz$- and $xz$-plane to reduce the model to quarter size in the domain $x>0$ and $y>0$ allowing for finer meshing and/or faster computations. We apply symmetric boundary conditions at the $yz$-plane $x=0$ and anti-symmetry at the $xz$-plane $y=0$,
\bsubalat{BC_symmetry}{3}
 \text{Symme}&\text{try, } x=0: &&&&
 \nn
 \\
 \partial_{x}\varphi &= 0,\;  &
 \pp_x p_\mr{1}^{\mr{fl}} &= 0,&
 \pp_x p_{2}^{\mr{fl}} &= 0,
 \\
 v_{1,x}^{\mr{sl}} &= 0,\; &
 \partial_{x} v_{1,y}^{\mr{sl}} &= 0,\quad  &
 \partial_{x} v_{1,z}^{\mr{sl}} &= 0,
 \\
 v_{2,x}^{\mr{fl}} &= 0,\; &
 \pp_x v_{2,y}^{\mr{fl}} &= 0,\; &
 \pp_x v_{2,z}^{\mr{fl}} &= 0,
 \\
 \text{Anti-symme}&\text{try, } y=0: &&&&
 \nn \\
 \varphi &= 0,\;  &
 p_\mr{1}^{\mr{fl}} &= 0,\; &
 \pp_y p_{2}^{\mr{fl}} &= 0,
 \\
 v_{1,x}^{\mr{sl}} &= 0,\; &
 \partial_{y}v_{1,y}^{\mr{sl}} &= 0,\; &
 v_{1,z}^{\mr{sl}} &= 0,
 \\
 v_{2,x}^{\mr{fl}} &= 0, &
 v_{2,y}^{\mr{fl}} &= 0,\; &
 \pp_y v_{2,z}^{\mr{fl}} &= 0.
 \esubalat

\section{Results comparing a thin-film and a bulk transducer device}
\seclab{ThinfilmVSbulk}

\subsection{The two main model devices}
\seclab{TwoMainDevices}

As a proof of concept that a tiny thin-film transducer is able to drive an acoustofluidic device as well as a conventional bulk transducer, we study the two devices shown in \figref{ThinfilmVSbulkPZT}(a,b), oriented along the $x$-, $y$- and $z$-axis, and both containing a water-filled microchannel of length $\Lwa = 35$~mm, width $\Wwa = 0.377$~mm, and height $\Hwa = 0.157$~mm, chosen as a typical channel size used in the literature\citep{Lenshof2012a} and specifically studied experimentally and theoretical in Ref.~\onlinecite{Muller2013}.

The thin-film device consist of a rectangular glass block of $\Lgl = 40$~mm, width $\Wgl = 3.02$~mm, and height $\Hgl = 1.4$~mm. An \AlScNiv\ thin-film transducer of length $\Lthf = \Lgl$, width $\Wthf = \Lgl$, and height $\Hthf=1~\SIum$, is deposited on the bottom surface of the $xy$-plane. The anti-symmetric voltage actuation is made possible by splitting the bottom electrode in half by a 40-$\SIum$-wide gap along the $x$-axis. Here, the microchannel is centered both along the $x$-axis and the $y$-axis, but its top  aligns with the center of the glass height $\frac12\Hgl$ to mimic that the glass block consists of two glass slabs of equal height bonded together, but with the microchannel etched into the top surface of the bottom slab. This specific device is chosen, because in a recent study, we successfully modeled and experimentally validated a similar thin-film-glass-block device without the microchannel.~\cite{Steckel2021} Note that the fraction of the total volume occupied by the thin-film transducer is minute, $\calV_\mr{thf}/\calV_\mr{tot} = 0.07\%$ v/v.

The bulk-transducer device has been studied extensively both experimentally and numerically in the literature.\citep{Barnkob2010, Augustsson2011, Muller2013} It consists of a silicon substrate of length $\Lsl = 40$~mm, width $\Wsl = 2.52$~mm, and height $\Hsi = 0.35$~mm, into which the centered microchannel is etched, and onto which is bonded a Pyrex glass lid of the same length and width, but with the height $\Hpy = 1.13$~mm. This silicon-glass chip is placed off-center on a nominal 2-MHz PZT-transducer of $\Lpzt = 40$~mm, width $\Wpzt = 5$~mm, and height $\Hpzt = 1$~mm, such that the right-most side walls align. In the actual device, the transducer is glued to chip, but here we neglect the glue layer and assume an ideal bonding instead. Note that the fraction of the total volume occupied by the bulk PZT transducer is large, $\calV_\mr{pzt}/\calV_\mr{tot} = 57\%$~v/v.

Using these dimensions and the material parameters listed in \tabsref{param_water}{param_solids} we implement these two 3D device models in the commercial finite-element software COMSOL Multiphysics 5.4,\citep{Comsol54} closely following the implementation method described in Ref.~\onlinecite{Skov2019}. All simulations are run on a workstation with a 16-core processor Intel i9-7960X @ $3.70\, \SIGHz$ boost clock and with 128 GB ram.

\begin{figure*}[!t]
\centering
\includegraphics[width=\textwidth]{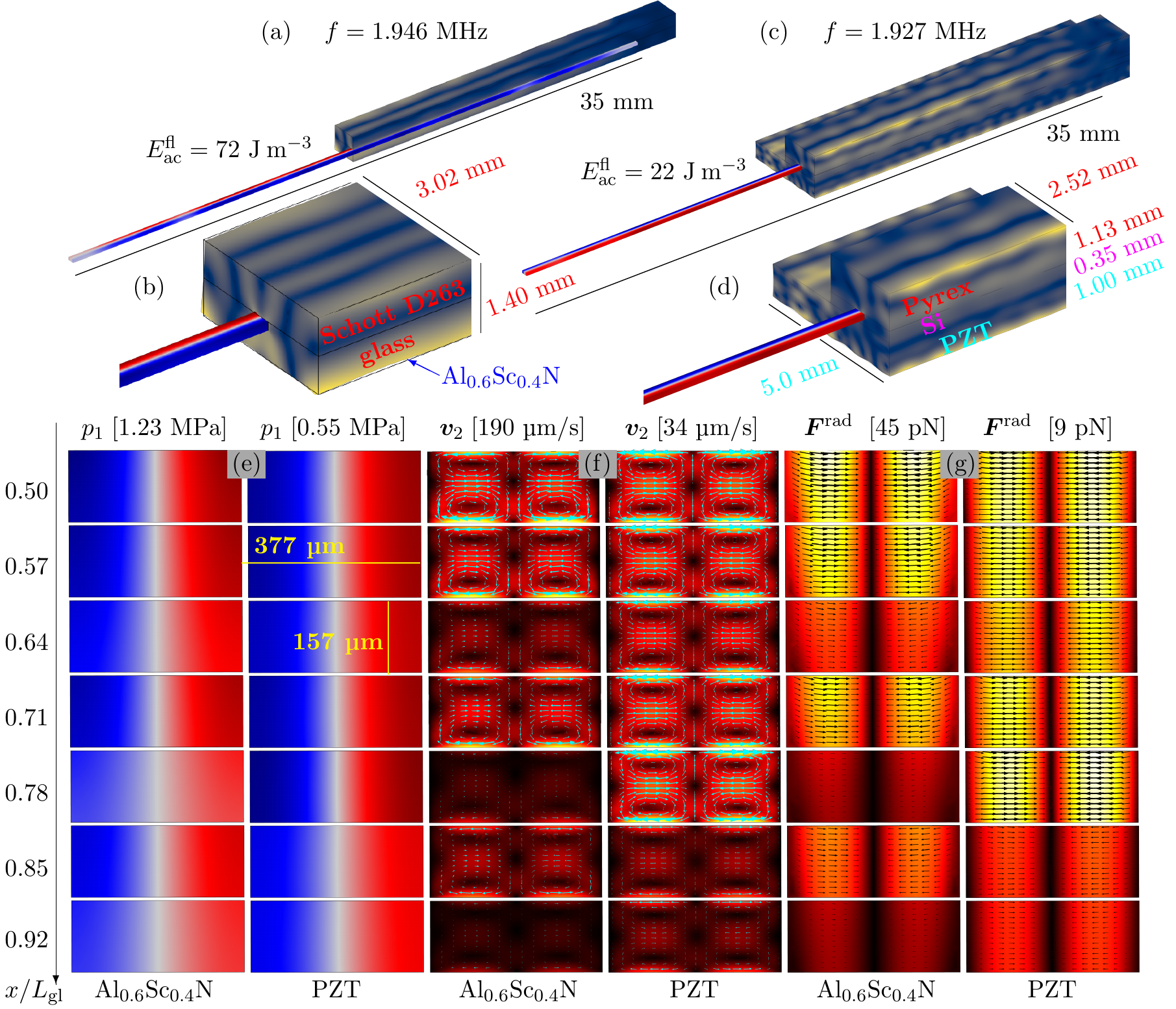}
\caption[]{\figlab{ThinfilmVSbulkPZT} (a,b) A glass chip driven by a 1-um-thick Al$_{0.6}$Sc$_{0.4}$N-thin-film transducer (not visible) at resonance$f = 1.946~\SIMHz$  actuated with $1~\mr{Vpp}$. The color plots show the displacement field $u_{1}$ from $0$ (blue) to $15~\SInm$ (yellow) and $p_1$ from $-1230~\SIkPa$ (blue) to $-1230~\SIkPa$ (red). (c,d) A conventional Si-glass chip driven by a bulk PZT at resonance $f = 1.927~\SIMHz$ actuated with $1~\mr{Vpp}$. The color plots show the displacement field $u$ from $0$ (blue) to $3.6~\SInm$ (yellow) and $p_1$ from $-550~\SIkPa$ (blue) to $+550~\SIkPa$(red). (d) Cross section plots for the bulk pressure field $p_{1}$ of the \AlScNiv driven system on the left and the PZT driven system of the right, with the lengths at which the cross sections were taken in the device. (e) Cross sections of the streaming velocity $v_2$ of the \AlScNiv driven system on the left and the PZT driven system of the right. (f) Cross sections of the radiation force $F^\mr{rad}$ for suspended 5-um-diameter polystyrene particles of the \AlScNiv driven system on the left and the PZT driven system of the right.}
\end{figure*}

\subsection{Mode analysis of the two devices}
\seclab{ModeAnalysis}

First step in our analysis is to identify good acoustic resonances in the two devices, which we actuate in a comparable manner with a peak-to-peak voltage of $\varphi_0 = 1$~V. In the thin-film device the voltage amplitude of the ac-voltage on the  "positive" ("negative") half of the top electrode is set to $+\frac12 \varphi_0$ ($-\frac12 \varphi_0$, 180$^\circ$ out of phase) relative to the grounded bottom electrode. Similarly, the voltage amplitude on the top electrode in the bulk-PZT device is set to $+\frac12 \varphi_0$ relative to the grounded bottom electrode. The frequency of the actuation voltage is then swept from 0.1 to 3.5~MHz, while monitoring the acoustic energy density $\Eacfl$, \eqref{EacDef}, in the water. The frequency steps in the sweep is adaptive, range from $\Delta f = 16$~kHz when the local curvature in $\Eacfl(f)$ is small (far from resonance peaks) down to $\Delta f = 0.03$~kHz when it is large (near resonance peaks).

As expected, the strongest resonance peak in $\Eacfl$ happens near the hard-ward standing half-wave resonance $f_0 = \frac{\cfl}{2\Wwa} = 2$~MHz. This main resonance is located at $\fthf = 1.946$~MHz with a maximum energy density of $\Eacfl(\fthf) = 72.1~\SIJpcm$ for the thin-film device, and at $\fpzt = 1.927$~MHz with a $\Eacfl(\fpzt) = 21.7~\SIJpcm$. The amplitude $p_1$ of the pressure and $\uuu_1$ of the displacement for these main resonance modes in the two devices are shown in \figref{ThinfilmVSbulkPZT}(b,c). One immediate conclusion is that the quality of the resulting resonant pressure mode in the two devices are comparable, a nearly perfect anti-symmetric wave across the channel with only weak variations along the channel, and the pressure amplitude of 1.23~MPa in the thin-film device 2.2 times the 0.55-MPa amplitude in the bulk-PZT device. Clearly, the tiny 0.07\% v/v thin-film transducer can deliver a fully comparable, and perhaps even better, acoustic response in the device in comparison with conventional large large 57\%~v/v bulk PZT transducer.

When inspecting the displacement field, it is seen that the displacement field has a more regular mode with a larger 15-nm amplitude in the thin-film device compared to the more complex resonance mode with a 3.6-nm amplitude in the larger volume of the bulk-PZT device. Perhaps this is an indication of the regular mode being more efficient in transferring acoustic energy from the transducer through the solid into the microchannel. In \figref{ThinfilmVSbulkPZT}(d) is shown the pressure in seven vertical channel cross sections equally spaced along the channel from its center to its end, showing the weak axial variations in $p_1$ for both devices. The bulk-PZT device appears more constant along the channel, however, it is 2.2 times weaker than $p_1$ in the thin-film device, in which pressure wave nevertheless dies out towards the ends of the channel.

\subsection{The acoustic radiation force and streaming at resonance}
\seclab{FradStreaming}

The acoustic modes $p_1$ and $\uuu_1$ are the basic fields giving rise to the steady time-averaged responses used for applications in acoustofluidic devices, namely the acoustic streaming $\vvv_2$ in the fluid and the radiation force $\FFFrad$ acting on suspended microparticles. In \figref{ThinfilmVSbulkPZT}(e,f) these responses are shown in the same seven cross sections as used in \figref{ThinfilmVSbulkPZT}(d). Being second-order repsonses, the improvement factor of 2.2 in pressure becomes an factor of $2.2^2 \approx 5$ in the streaming and radiation force. The axial variations are like wise augmented, but the overall conclusion remains the same: the time-averaged response of the thin-film device is the same as that of the bulk-PZT device. It supports the usual quadrupolar Rayleigh streaming pattern, and the radiation force points towards the vertical center plane along the axis, which can thus serve as a plan for particle focusing. The focusing force is nearly 5 times larger in the thin-film device compared to the bulk-PZT device, and from this first example it is clear that the thin-film device would work as an excellent acoustofluidic device if manufactured. In the following we investigate further the characteristics of the thin-film device.

\section{Physical aspects of acoustofluidic devices driven by thin-film transducers}
\seclab{PhysicalAspects}

In this section we use the numerical model to study various physical aspect of acoustofluidic devices with thin-film transducers. The study includes the physical principle of the thin-film transduction process, the robustness of the device to material, thickness, quality factor of the thin-film transducer, the role of the shape of the thin-film-transducer electrodes, and sensitivity of the device to shifts in the channel position away from exact centering in the glass chip.

\subsection{The physical mechanism of thin-film-actuated bulk acoustic waves}
\seclab{PhysicalMechanism}

The ability of the thin-film transducer to create the desired acoustofluidic effects in a bulk acoustofluidic device rely on three physical aspects of the system: The in-plane-expansion of the the thin-film transducer under the action of the orthogonal applied electric field, the acoustic whole-system resonance of the device, and the high Q-factor of the elastic solid constituting the bulk part of the device.

Traditional bulk PZE transducers typically work by exciting a standing half-wave in the mechanical displacement field along the polarization axis of the transducer by applying an electric field in the same direction, facilitated by a large longitudinal PZE coefficient $e_{33}$. This thickness mode is fairly easy to compute and to design for. Moreover, this mode is also relatively good at transferring acoustical power into a device attached to the transducer, by the large component along the surface normal of the induced displacement field. Conventional bulk PZE transducers with a millimeter-thickness typically have good resonances in the low MHz regime.

\begin{figure}[!t]
 \centering
 \includegraphics[width=\columnwidth]{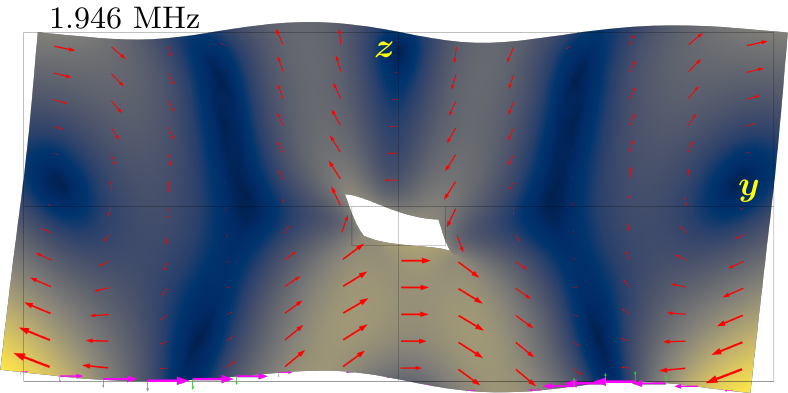}
 \caption[]{\figlab{FilmExpansion} Numerical simulation of displacement field $\uuu_1$ in the \AlScNiv\ thin-film device of \figref{ThinfilmVSbulkPZT}(a), enhanced by a factor 7000 for clarity, of the whole-system resonance $\fthf = 1.946$~MHz. Vector plot of  $\uuu_1$ (red vectors) and color plot of its magnitude $|\uuu_1|$ from 0 (dark blue) to 15~nm (light yellow). The mode is excited by the \AlScNiv\ thin-film transducer driven by a 1-\Vpp\ AC-voltage at the frequency $f = \fthf = 1.946$~MHz. The in-plane strain $\pp_y u_{1,y}\:\een_y$ (magenta vectors) generated by the transducer on the transducer-glass interface showing an expansion (contraction) on the left (right) side compatible with strain of whole-system resonance mode.}
 \end{figure}

In contrast, the half-wave transducer resonances of the thin-film transducers are pushed up into the GHz regime due to the micrometer-sized thickness, much higher than the low-MHz frequencies usually used in acoustofluidics. Therefore, the transduction studied in this work is dominated by the transverse PZE coefficient $e_{31}$. The large electric field in the order of MV/m that results from applying, say, a potential difference of 1~\Vpp\ across a 1-$\SImum$-thick thin-film transducer, generates a large strain that accumulates along the millimeter-sized transducer-glass interface, the first of the three physical prerequisites. When this strain oscillates at a resonance frequency of the glass block constituting bulk part of the device, the corresponding eigenmode of the glass block is excited if it has a compatible strain pattern, the second physical prerequisite. We emphasize that this transduction mechanism does not rely on exciting any resonances in the thin-film transducer, but instead on exciting resonances in the whole system, of which the transducer is only a minute part.

The thin-film transduction mechanism is exemplified in in \figref{FilmExpansion} by the mode $\fthf$ the thin-film device \figref{ThinfilmVSbulkPZT}(a). The in-plane strain $\pp_y u_{1,y}\:\een_y$ on the transducer-glass interface generated by the anti-symmetrically driven split-electrode thin-film transducer correspond to an expansion on the left side and a contraction on the right side. This strain-pattern is compatible with that of the whole-system resonance mode, which therefore is excited with a large 15-nm displacement amplitude. The resulting anti-symmetric oscillatory displacement field of the glass block pushes on the water in the channel, which leads an anti-symmetric pressure wave $p_1$, \figref{ThinfilmVSbulkPZT}(ed), with the desired acoustofluidic properties shown in \figref{ThinfilmVSbulkPZT}(e-f). The third and last physical prerequisite is the high quality factor of the whole system as an acoustic resonator. Because the thin-film transducer in our main example constitutes only a 0.07\% v/v, the quality factor is completely dominated by that of the glass block, which has a high value $Q\sim 10^3$ in a typical glass.\citep{Hahn2015} Note that the pressure wave in the water-filled microchannel does not need to be a resonant standing half-wave, but if it is, its amplitude may be enhanced further.

\begin{figure}[!t]
 \centering
 \includegraphics[width=0.9\columnwidth]{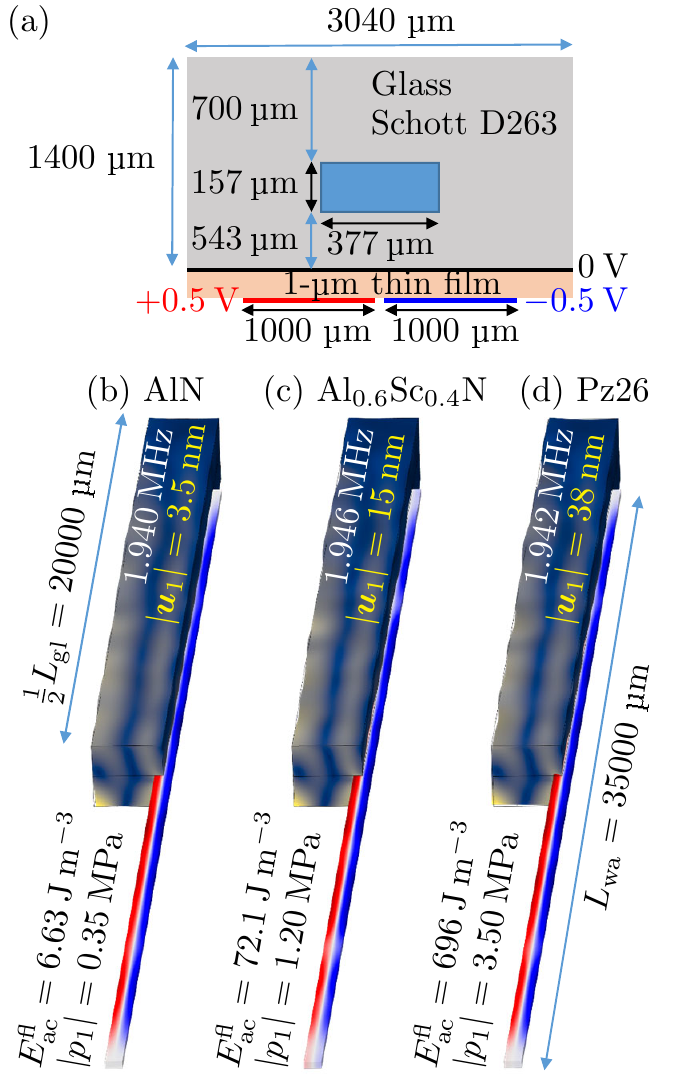}
 \caption[]{\figlab{PZEmaterials} (a) Sketch of cross section of the thin-film driven system used the simulations shown on (b)-(d), where the two electrodes are actuated out of phase, with a gap of $40~\SIum$ between them. The glass, of type Schott D263, devices are $40000~\SIum$ long and the water channel is only $35000~\SIum$, so the last part the channel is closed off by the glass. (b) Simulation of a $1~\SIum$ thick AlN thin-film, (c) simulation of a $1~\SIum$ thick Al$_{0.6}$Sc$_{0.4}$N thin-film, and (d) simulation of a $1~\SIum$ thick Pz26 thin-film. The Pz26 parameters are for the purpose of the simulations assumed to have the same parameters as bulk Pz26. The different films show that they all have the same mode, with a slight shift in frequency, and for $1~\mr{Vpp}$ Pz26 has the largest resonce, followed by Al$_{0.6}$Sc$_{0.4}$N and then AlN, although the breakdown voltage is also different for the different films.}
 \end{figure}

\subsection{The robustness of the device to material, thickness, and quality factor of the thin-film transducer}
\seclab{ThinfilmRobustness}

The above-mentioned thin-film transduction method implies that the functionality of the thin-film acoustofluidic device is robust to changes in several characteristic properties of the thin-film transducer, essentially because of its small volume fraction of the whole system. In the following we study three of such properties,  namely the material, thickness, and quality factor of the thin-film transducer.

We study three types of PZE materials. One is the commonly used and commercially available PZT having large PZE coefficient $e_{33}$. One drawback of this material is its lead contents, which is being out-phased for health- and environmental reasons, and another is the difficulty in fabricating the material with a sufficiently low dissipation. Whereas other materials have lower PZE coefficients than PZT, they may be fabricated with higher purity and less dissipation. The lead-free AlN is a choice for its simpler and more well-controlled high-quality depositing on a variety of substrates, which allows for higher break-down voltages that almost make op for the lower PZE coefficient. \AlScNx\ offers a PZE coefficient between the values of PZT and AlN, with many of the same advantages as AlN, but has a more complex fabrication process and a lower break-down voltage. In \figref{PZEmaterials} we show simulation results three different PZE materials, while keep all other quantities fixed in the model: AlN, \AlScNiv, and PZT Pz26, maintaining this order when referring to the numerical results in the following. In spite of the very different material parameters listed in \tabref{param_solids}, the resulting acoustofluidic response of the main resonance $\fthf$ is nearly the same. Within 0.3\%,  the resonance frequencies are identical $\fthf = 1.940$, 1.946, and 1.942~Mz, where as the field amplitudes reflect the difference in the PZE coefficients and are for the pressure $|p_1| = 0.35$, 1.20, and 3.50~MPa, the displacement $|\uuu_1| = 3.5$, 15, and 38~nm, and the acoustic energy density $\Eacfl = 6.66$, 72.1, and 669~$\SIJpcm$. Besides the obvious difference in amplitude, the computed whole-system resonance is nearly the same in all three cases, showing a nearly ideal anti-symmetric pressure wave in the microchannel.

\begin{figure}[t]
 \centering
 \includegraphics[width=\columnwidth]{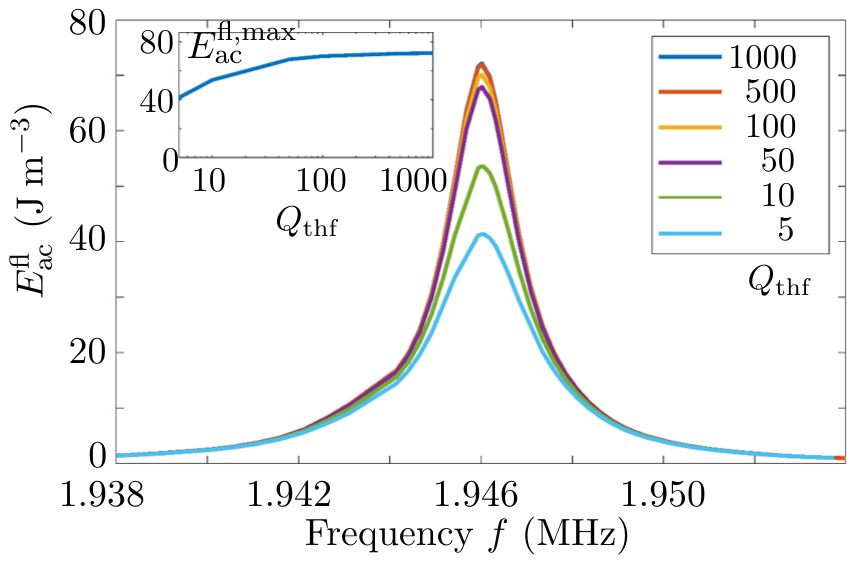}
 \caption[]{\figlab{FilmQuality} The acoustic energy density $\Eacfl(f)$ of a resonance peak for the system shown in \figref{ThinfilmVSbulkPZT}(a) for $f = 1.938$ to $1.954~\SIMHz$ for the thin-film Q-factor $Q_\mr{thf}$ in the range from 1000 to 5. The insert is a plot of the maximum $\Eacfl$ vs.~$Q_\mr{thf}$.}
 \end{figure}

The thin-film device is also insensitive to the quality factor of the thin-film transducer, which in terms of the damping coefficient $\Gamsl$ in the Cauchy equation~\eqnoref{CauchyEq}, is given by $Q = \frac{1}{2\Gamsl}$. For two reasons, we expect a weak dependency on $Q$. The smallness of the transducer implies that the Q-factor of the system is completely dominated by that of the glass block, and since the transduction mechanism does not rely on resonance properties of the thin-film transducer, the strong $Q$-dependence usually associated with resonant modes is absent. The simulation results shown in \figref{FilmQuality} confirms our expectation. Here, the acoustic energy density $\Eacfl$ in the microchannel of the thin-film device \figref{ThinfilmVSbulkPZT}(a) at the resonance $\fthf = 1.946$~MHz is shown as a function of $Q$ from the original value of 1000 down to an appalling low value of 5. The resonant behavior reflected in $\Eacfl(f)$ is maintained, and the change in $Q$ by a factor of 200 results in a drop of the peak value of $\Eacfl$ of 2, from 75 to $40~\SIJpcm$.

Finally, in \figref{Film_thickness} the simulation result shows that the main resonance mode $\fthf$ is maintained when changing the thickness by a factor of 3 from $\Hthf = 1$ to $3~\SImum$, numbers typical for current AlN and \AlScNx\ thin-film fabrication technology. The amplitude of the resonance peak in $\Eacfl$ decreases from 6.9 to $6.6~\SIJpcm$ from thinnest to the thickest film.

\begin{figure}[!t]
 \centering
 \includegraphics[width=\columnwidth]{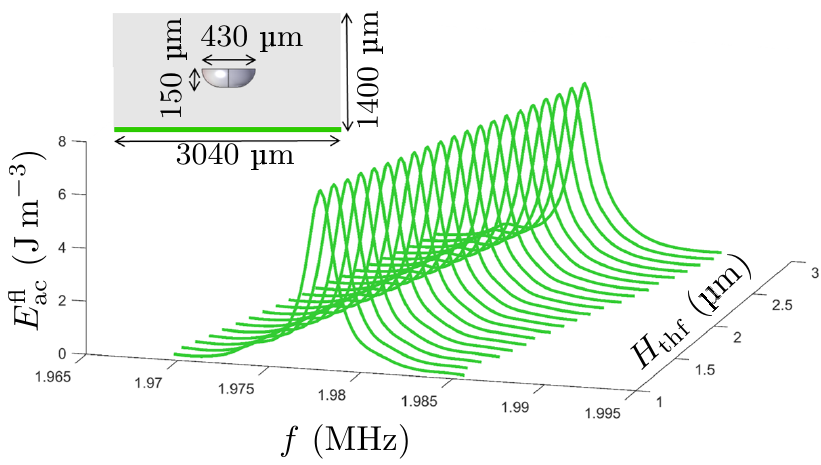}
 \caption[]{\figlab{Film_thickness} 3D simulations of $\Eacfl(f)$ in a Schott D263 glass device with an AlN thin-film transducer thickness $\Hthf$. The main resonance peak in $\Eacfl$ is shifted from $\fthf = 1.977$ to $1.985~\SIMHz$ as $\Hthf$ is increased from 1 to 3$~\SIum$, and the corresponding maximum $\Eacfl$ is decreased from 6.9 to $6.5~\SIPa$. The inset shows the cross section of the 40-mm-long device. The microchannel has semi-circular side walls mimicking isotropic etching in glass.
 }
 \end{figure}

\subsection{Enhancing the acoustic response of the device by shaping the electrodes of the thin-film transducer}
\seclab{ElectrodeShape}

Thin-film transducers are fabricated by standard microfabrication deposition techniques, and this implies several distinct advantages. The lateral shape of the transducer or its electrode can be chosen freely by photolithography techniques, the attachment of the transducer to the glass chip is reproducible, stable and strong, and the less controlled use of glue known form standard bulk transducer technology is avoided. Commercially, microfabrication techniques open up for volume production with relatively cheap unit prices, a necessary prerequisite for widespread single-use applications in biotechnology and medicine, where the cross-contamination arising from multiple use of the same device is a no go.

An illustrative example of how the shape of the metal electrodes on the surface of the thin-film transducer may enhance the acoustic response, is shown in \figref{Include_electrodes}. Here we show a pyrex-glass block of width $\Wsl = 2.8$~mm and height $\Hsl = 1.4$~mm containing a rectangular channel of  $\Wwa = 0.4$~mm and height $\Hsl = 0.15$~mm. A 2-$\SImum$-thick AlN thin-film transducer is attached to the top surface. Two electrode configurations are considered. In \figref{Include_electrodes}(a), the top and bottom electrodes cover the entire thin-film surface, and therefore even resonance modes can be excited having $F(x,y,z) = F(x,-y,z)$ for any field $F$. A strong whole-system resonance is located at $\fthf = 3.49$~MHz, where the associated pressure wave of magnitude $|p_1| = 24$~kPa in the channel have two vertical nodal planes (gray) place symmetrically around the center plane $y=0$, in contrast to the anti-symmetric mode with a single vertical nodal plane at the center $y=0$ shown in \figref{ThinfilmVSbulkPZT}(d).

In \figref{Include_electrodes}(b) the same device is shown, but now with the middle half of the top electrode removed. Of course, given this minute change in the system, the same whole-system resonance mode is excited, but the \textit{diminished} electrode coverage has lead to an \textit{increase} in the pressure amplitude by a factor of 6 to 146~kPa. The explanation of this perhaps surprising result is found in the spatial form of the whole-system resonance mode. By inspection we see that the displacement field at the glass-transducer interface forms a wave with in-plane contractions and expansions. The PZE coefficient $e_{31}$ in the transducer implies the presence of an electric field with a vertical component that changes sign along the in-plane direction. This tendency is counteracted by the fully covering top electrode, which imposed a unidirectional electrical field. Consequently, by removing the central part of the top electrode, this constraining boundary condition is relaxed, while the remaining side parts of the electrode are still capable of exciting the whole-system resonance mode. This example offers a glimpse of the opportunities for design improvements by performing a shape optimization of the electrodes or perhaps the entire thin-film transducer.

\begin{figure}[!t]
 \centering
 \includegraphics[width=0.8\columnwidth]{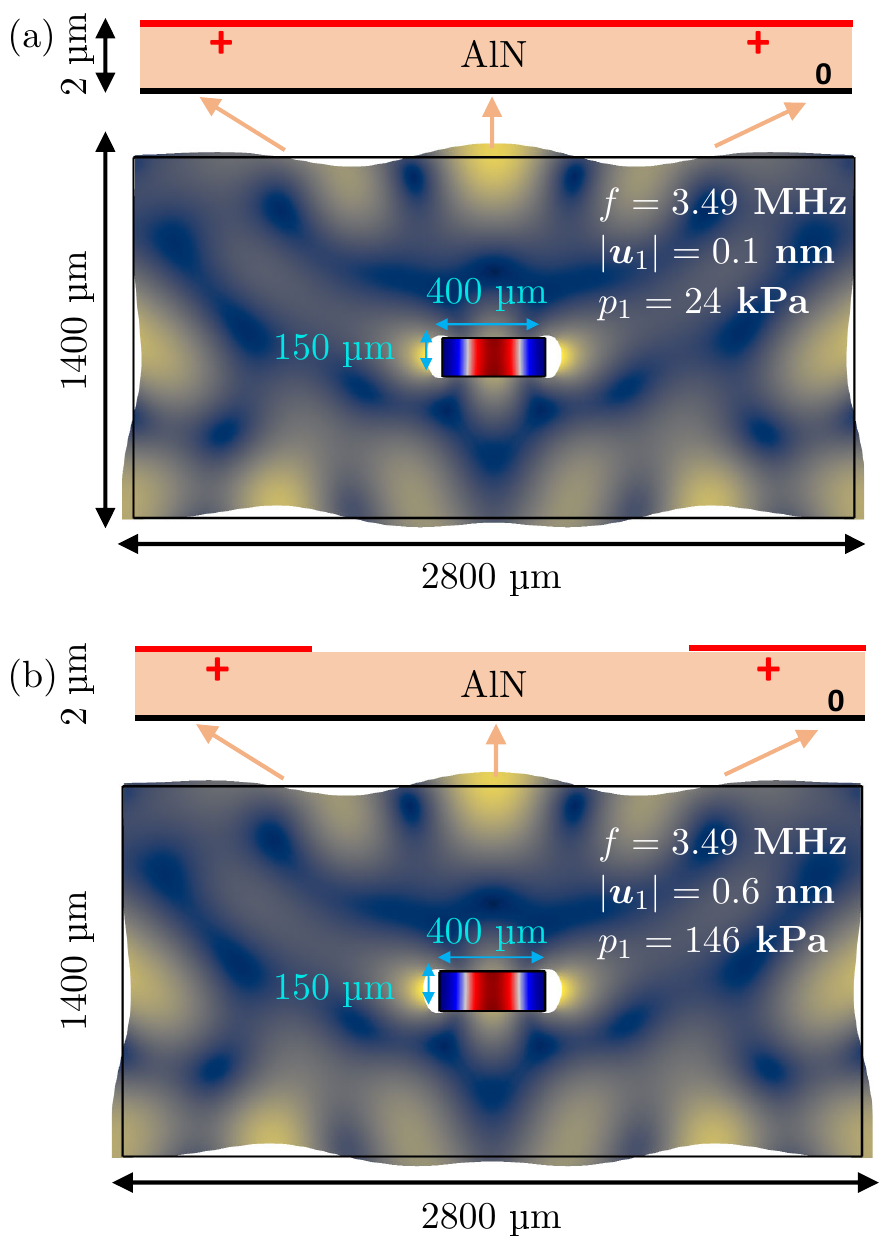}
 \caption[]{\figlab{Include_electrodes}  (a) 2D simulations of 2-$\SIum$-AlN thin-film transducer actuated by a un-split top electrode for symmetric actuation at 1$~\mr{V_{pp}}$, with a frequency $f=3.49~\SIMHz$ that actuates the standing second nodal standing wave in the horizontal direction in the liquid, with a pressure amplitude of $p_{1}=\pm~24~\SIkPa$. The device is 2800$~\SIum$-wide and 1400$~\SIum$ thick, and the substrate material is Pyrex. (b) same dimensions, materials, and frequency as in (a) but with half of the electrode cut away, which gives a maximum pressure amplitude of $p_{1}=\pm~146~\SIkPa$.}
 \end{figure}

\begin{figure}[!t]
 \centering
 \includegraphics[width=\columnwidth]{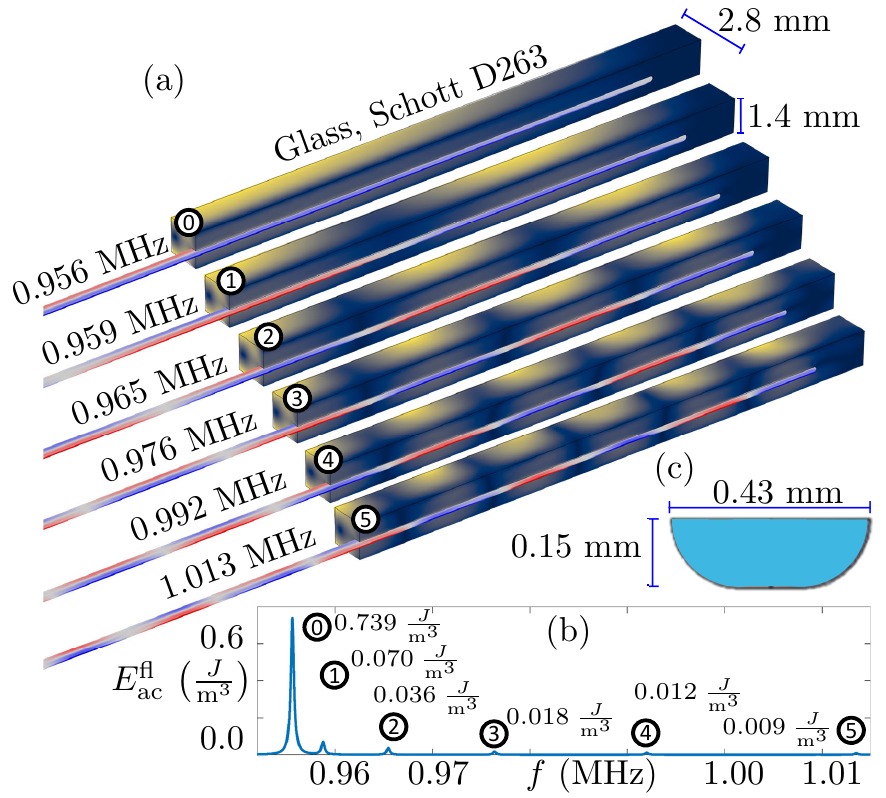}
 \caption[]{\figlab{CleanModes} Simulation of the low-frequency modes near 1~MHz in a thin-film device with a 1-$\SImum$-thick split bottom-electrode AlN thin-film transducer. The system is symmetric around the vertical center plane at $x=0$ and anti-symmetric around the vertical center plane at $y=0$. (a) The lowest six resonance pressure modes $n = 0, 1, \ldots, 5$ with $2n$ nodal planes along the $x$ axis, and 1 nodal plane along $y$. (b) The acoustic energy density spectrum $\Eacfl(f)$ identifying the six resonance modes. (c) The cross-sectional shape of the microchannel in the vertical $yz$-plane.}
 \end{figure}

\subsection{Spatially regular modes in the thin-film device}
\seclab{regular_modes}

Intuitively, the simplicity of the thin-film device consisting essentially of just a glass block should lead to simpler modes with regular spatial dependencies. As mentioned above, the presence of a bulky PZT transducer leads to the excitation of whole-system resonance modes with a more irregular wave pattern in the displacement fields. Also experimentally, this has been observed as hot spots in the pressure field along an otherwise perfectly shaped rectangular microchannel.\citep{Augustsson2011}

In \figref{CleanModes} we show the lowest whole-system resonance modes in a thin-film device with a 1-$\SImum$-thick split top-electrode AlN thin-film transducer mounted on the bottom of a rectangular Schott D263 glass block of length $\Lsl = 45$~mm, width $\Wsl = 2.8$~mm, and height $\Hsl = 1.4$~mm. The microchannel of the system has length $\Lwa = 40$~mm, width (at its top) $\Wwa = 0.4$~mm, and height $\Hwa = 0.15$~mm. To mimic the shape obtained by isotropic etching in glass, the side walls are modeled as quarter circles. The anti-symmetric actuation of the split top electrode combined with the geometrical symmetry dictates that the system is symmetric around the vertical plane at $x=0$ across the device and anti-symmetric around the vertical plane at $y=0$ along the device. In \figref{CleanModes}(a), one immediately notice the spatial regularity of both the displacement field $\uuu_1$ and the pressure field $p_1$ in the six displayed modes. Both fields have the required symmetry along the $x$-axis and anti-symmetry along the $y$-axis, and both fields exhibits one nodal plane along the transverse $y$ direction and respectively $2n$ nodal planes with $n = 0, 1, 2, \ldots , 5$ along the axial $x$ direction. In \figref{CleanModes}(b) is shown the spectrum $\Eacfl(f)$ in the frequency range from 0.952 to 1.015~MHz, which allows the identification of the six resonance frequencies $f_{n,1,0}$, where the indices refer to the number of nodal planes in each direction. Of the six modes, the $n=0$-mode without nodes along the $x$-axis has an axial structure that matches the $x$-independent voltage boundary condition better than the other modes, and indeed it has the highest energy density. As the number $n$ of $x$-axis nodes increases, the corresponding mode exhibits an increasing number of nodes, and thus an increasing mismatch with the $x$-independent voltage boundary condition. This explains the monotonically decreasing peak value of $\Eacfl$ for increasing values of $n$ seen in \figref{CleanModes}(b).

\begin{figure}[b]
 \centering
 \includegraphics[width=\columnwidth]{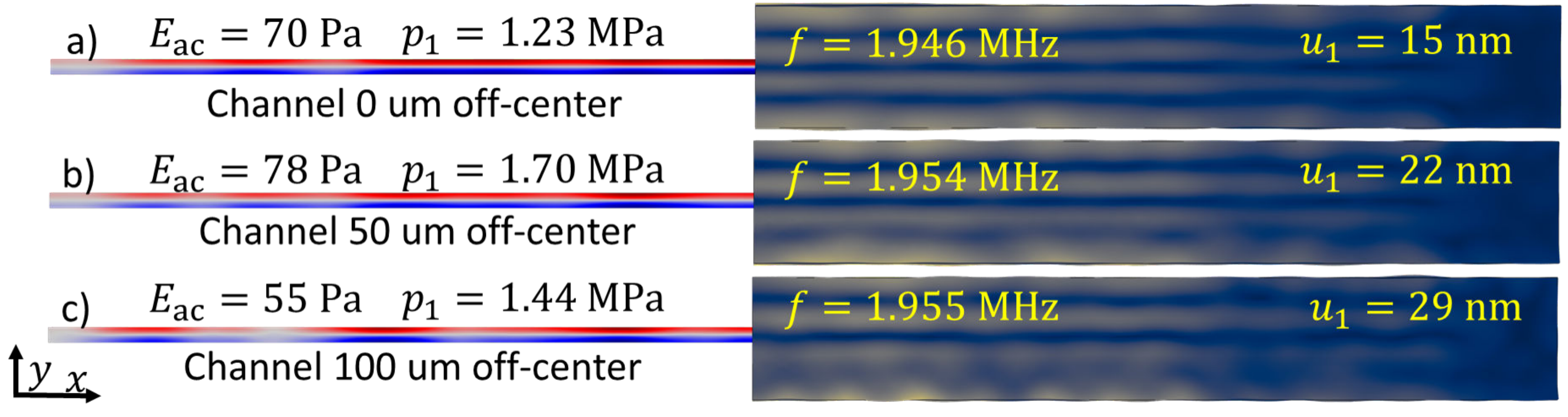}
 \caption[]{\figlab{Resonance_Electrode_channel_off-set} (a) Acoustic resonances of the half system of same dimensions as in \figref{ThinfilmVSbulkPZT}(a) for a frequency of $f=1.954~\SIMHz$, an acoustical energy of $E_{ac}= 70~\SIPa$, a maximum pressure amplitude $p_1=\pm 1.23~\SIMPa$, and a max displacement of $u_{1} = 15~\SInm$. (b) Acoustic resonance at $1.954~\SIMHz$ of the same glass dimensions as in (a), but with the channel off-set by 50$~\SIum$, with an acoustical energy of $E_{ac}= 78~\SIPa$, a maximum pressure amplitude $p_1=\pm 1.70~\SIMPa$, and a max displacement of $u_{1} = 22~\SInm$. (c) Acoustic resonance at $1.955~\SIMHz$ where the glass dimensions are unchanged but the channel is off-set with 100$~\SIum$ and it has an acoustical energy of $E_{ac}= 55~\SIPa$, a maximum pressure amplitude $p_1=\pm 1.44~\SIPa$, and a max displacement of $u_{1} = 29~\SInm$.}
\end{figure}

\subsection{Device performance as a function of breaking the geometrical symmetry}
\seclab{SymmetryBreaking}

As a final point, we discuss the consequences of breaking the perfect anti-symmetry of the thin-film device imposed in Figs.~\fignoref{ThinfilmVSbulkPZT}, \fignoref{FilmExpansion}, \fignoref{PZEmaterials}, and~\fignoref{CleanModes}. Using microfabrication techniques, many geometrical features can be defined with accuracies down between 1 and 10~$\SImum$, however it can be problematic to reach such accuracies when dicing up a full-sized wafer into the individual devices. For microelectronics this is not problematic, if the integrated circuits is sufficiently removed from the edges. However, for acoustofluidic devices the whole substrate influences the whole-system resonances. For this reason it is interesting to investigate the sensitivity of a given acoustofluidic device given shifts in the position of the microchannels relative to the edges of the substrate.

In \figref{Resonance_Electrode_channel_off-set}, we study the acoustic response to a shift in the center axis of the microchannel in the thin-film device of \figref{ThinfilmVSbulkPZT}(a) from the ideal (anti-)symmetric position at $y = 0$ to 50 and 100~$\SImum$. It is gratifying to see that the whole system resonance mode is not degraded significantly. A contributing factor to this robustness is that the water-filled channel only constitutes 1.2\% of the total volume of the device. We notice that the main anti-symmetric form of the acoustic pressure is unaffected by the shift, and the acoustic energy density $\Eacfl$ remains high, in the range from 55 to 70~$\SIJpcm$. However, as the shift increases, more pronounced axial inhomogeneities develop. For application in the stop-flow mode this could imply a degradation in functionality, however as is well known experimentally in several acoustofluidic devices, in flow-through applications such axial inhomogeneities averages out, and the device would work essentially without degradation.\citep{Lenshof2012}

\section{Conclusion}
\seclab{Conclusion}

In this paper, based on a  well-tested and experimentally validated numerical model,\citep{Skov2019, Skov2019b, Bode2020} we have by 3D numerical simulations shown in \figref{ThinfilmVSbulkPZT} that glass chips with integrated piezoelectric thin-film transducers constituting less than 0.1\% v/v of the system, have an acoustofluidic response fully on par with that obtained in a conventional silicon-glass device actuated by a bulk lead-zirconate-titanate (PZT) transducer.  In \secref{PhysicalAspects}, we have shown that the ability of the thin-film transducer to create the desired acoustic effects in a bulk acoustofluidic device relies on three physical aspects of the system: The in-plane-expansion of the the thin-film transducer under the action of the orthogonal applied electric field, the acoustic whole-system resonances of the device, and the high Q-factor of the elastic solid constituting the bulk part of the device.

There are several advantages to the use of thin-film transducers. Among them are the low sensitivity of the thin-film device to the material, the thickness, and the quality factor of the thin-film transducer discussed in \secref{ThinfilmRobustness}. It is also an advantage that thin-film devices can be produced by clean-room microfabrication technique, similar to the ones employed in the fabrication of surface acoustic waves. An example of this was studied in \secref{ElectrodeShape}, where we demonstrated an enhanced acoustofluidic response by shaping the electrodes of the thin-film transducer, in this case by removing roughly half of the top-electrode coverage. Other advantages of microfabrication techniques are that the thin-film transducers are integrated in the devices in a much more reproducible manner compared to the conventional bulk-transducer technique involving the use of glue.

In an application perspective, the use of thin-film transducers offers new possibilities in the field of acoustofluidcs. The fact that the thin-film transducer constitutes such a low volume fraction implies not only that the device is relatively insensitive to the quality of the thin film, but also that the core part of the acoustofludic system, namely the microchannel, constitutes a relatively large part of the system and is thus easier to control. Several of the thin-film transducers can be fabricated with a high breakdown voltages ($\sim 20~\SIV/\SImum$) that allows for relatively high acoustic energy densities and lower dissipation and heat production.

We hope that this theoretical analysis will inspire our experimental colleagues in the field to investigate the new application aspects offered by the thin-film acoustofluidic devices.

\section*{Acknowledgements}

This work was supported by the \textit{BioWings} project funded by the European Union's Horizon 2020 \textit{Future and Emerging Technologies} (FET) programme, grant No.~801267.

%
%


\end{document}